\documentclass[%
 reprint,
 amsmath,amssymb,
 aps,
]{revtex4-2}

\usepackage{graphicx}
\usepackage{dcolumn}
\usepackage{bm}
\usepackage{braket}
\usepackage{hyperref}

\begin{document}

\preprint{APS/123-QED}

\title{Polaritons under Extensive Disordered Molecular Rotation in Optical Cavities}

\author{Wei Liu}
\affiliation{Department of Chemistry, School of Science, Westlake University, Hangzhou 310024 Zhejiang, China}
\affiliation{Institute of Natural Sciences, Westlake Institute for Advanced Study, Hangzhou 310024 Zhejiang, China}
\author{Jingqi Chen}%
\affiliation{Department of Chemistry, School of Science, Westlake University, Hangzhou 310024 Zhejiang, China}
\affiliation{Institute of Natural Sciences, Westlake Institute for Advanced Study, Hangzhou 310024 Zhejiang, China}

\author{Wenjie Dou}
\email{douwenjie@westlake.edu.cn}
\affiliation{Department of Chemistry, School of Science, Westlake University, Hangzhou 310024 Zhejiang, China}
\affiliation{Department of Physics, School of Science, Westlake University, Hangzhou 310024 Zhejiang, China}
\affiliation{Institute of Natural Sciences, Westlake Institute for Advanced Study, Hangzhou 310024 Zhejiang, China}

\date{\today}

\begin{abstract}
This study investigates the dynamic behavior of polaritons in an optical cavity containing one million molecules, emphasizing the influence of molecular rotation and level disorder on the coupling between molecules and photons. Through rigorous theoretical simulations and numerical analyses, we systematically explore the formation and spectral characteristics of polaritons in this complex environment. Our findings reveal that the rotational motion of molecules significantly affects the electromagnetic field distribution within the cavity, leading to distinct alterations in polariton properties. Simultaneously, the presence of level disorder induces diverse energy level structures, influencing the energy distribution of polaritons. The comprehensive examination of these factors provides valuable insights into the intricate interplay between molecules and photons in large-scale cavity systems. This research not only advances the fundamental understanding of molecular-photon coupling but also offers theoretical guidance for practical applications in the design and exploration of optical cavities.
\end{abstract}

\maketitle

\textit{Introduction.}—The interaction between matter (atoms, molecules, or solid materials) and the quantized electromagnetic field inside an optical cavity gives rise to a distinct set of states known as polariton states \cite{nagarajan2021chemistry, garcia2021manipulating, yuen2019polariton, hutchison2012modifying, ribeiro2021enhanced,xiang2018two, wang2021light, ribeiro2022multimode,welakuh2023tunable}. These states involve shared excitations between molecules and the cavity, influencing chemical reactivity \cite{nagarajan2021chemistry, hutchison2012modifying, thomas2019tilting, kowalewski2016cavity, kowalewski2016non, galego2016suppressing, herrera2016cavity, ribeiro2018polariton, feist2018polaritonic, lacombe2019exact, campos2019resonant, mandal2020polariton, mandal2020polariton, du2018theory, campos2019resonant, mandal2019investigating, li2021theory,li2021cavity,pannir2022driving,mandal2022theory,mandal2019investigating}. Describing them comprehensively is challenging due to the numerous light modes in the cavity and energetic, spatial, and orientational disorder.

Many theories simplify light by using a single cavity mode coupled to numerous quantum emitters \cite{chavez2021disorder, engelhardt2022unusual, feist2015extraordinary, sommer2021molecular, spano2015optical, shammah2017superradiance, botzung2020dark, dubail2022large, zhang2021collective, cohn2022vibrational, herrera2016cavity, houdre1996vacuum, xiang2019state, reitz2018energy, schafer2019modification, xiang2019manipulating, cao2022generalized, sun2022accurate,cui2022collective, finkelstein2023non, zhang2023multidimensional,mandal2023theoretical}. Recent research predicts changes in transport and relaxation based on disorder but often neglects practical factors like molecular rotation \cite{chavez2021disorder, engelhardt2022unusual}. The impact of molecular rotation on polaritons, especially in cavities with numerous molecules, remains unclear.
As molecules rotate, the dynamic nature of light-matter coupling, exhibiting time-dependent oscillations, comes to the forefront. The periodically driven Hamiltonian has been studied using Floquet theory \cite{chen2023floquet, liu2023optical, wang2023nonadiabatic, batge2023periodically}, and disorder's impact on polaritons has been explored \cite{izrailev1998classical, agranovich2003cavity, litinskaya2006loss, litinskaya2004fast, sun2022dynamics, engelhardt2022unusual, engelhardt2023polariton,ribeiro2023vibrational}. However, a comprehensive understanding of the dynamics of light and matter for quantum emitters driven by both Floquet and disorder is still lacking.

In this letter, we delve into a large cavity system comprising $10^6$ molecules and one photon, considering molecular rotation and disorder. Through theoretical simulations and calculations, we investigate the dynamic behavior and spectral features of polaritons in this intricate setting. The introduction of molecular rotation and disorder induces shifts in the energy levels of polaritons, impacting the oscillation period of photon-matter hybrid states. Increasing disorder leads to a rapid decay of these hybrid states, almost causing their disappearance. In the case of weak coupling, even though polaritons are not generated, molecular rotation is observed to extend the lifetime of the photon state.

\begin{center}
\begin{figure}
    \centering
    \includegraphics[width=.98\textwidth]{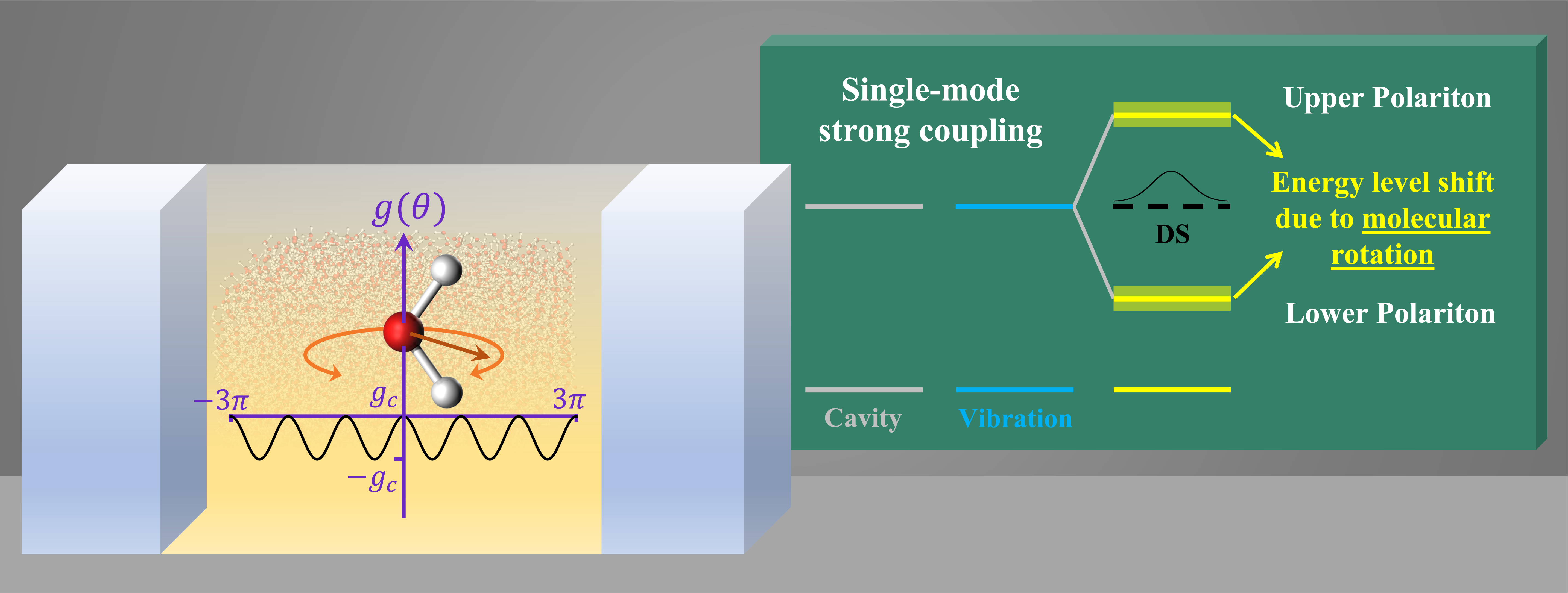}
    \caption{(a) Schematic displays that many molecules ($10^{6}$) are contained in a microcavity formed by two nearly $100 \%$ reflecting mirrors, considering the rotation of molecules. The intensity of light-matter interactions, $g$, fluctuates within the range of $[-g_{\text{c}}, g_{\text{c}}]$ with the rotation angle $\theta$ of the molecule. When a cavity resonance is tuned to match a molecular transition, it is possible to establish strong coupling, leading to two new “polariton” states , denoted as upper polariton (UP) and lower polariton (LP), split in energy proportionally to the coupling. The energy levels of these polaritons can be shifted by different molecular rotational angular velocities. The Gaussian distribution of dark states (DS) represents energy level disorder. }
    \label{fig:schematic}
\end{figure}
\end{center}

\textit{Floquet Disordered Tavis-Cummings Model.}—As shown in Fig.~\ref{fig:schematic}, we consider that many molecules ($10^6$) are contained in a microcavity formed by two nearly $100 \%$ reflecting mirrors, involving the rotation of molecules. We adopt a singlemode Floquet-driven disordered Tavis-Cummings model, whose Hamiltonian is given as $\hat{H}=\hat{H}_{\text{M}}+\hat{H}_{\text{C}}+\hat{H}_{\text{MC}}$, where
\begin{equation}
\begin{split}
    \hat{H}_{\text{M}} &= \sum_{j=1}^{N_{\text{M}}} \hbar \omega_{j}\hat{B}_{j}^{\dagger}\hat{B}_{j}, \
    \hat{H}_{\text{C}} = \hbar \omega_{\text{c}} \hat{a}^{\dagger}\hat{a}, \\
    \hat{H}_{\text{MC}} &= \sum_{j=1}^{N_{\text{M}}} g_{\text{c}} \cos(\Omega t+\phi)\hat{B}_{j}^{\dagger}\hat{a}+\text{H.c.} \label{eq:final}
\end{split}
\end{equation}
The molecules $j$ are described by bosonic operators $\hat{B}_j$. Here, the excitation energies $\hbar \omega_{j}$ are distributed according to a Gaussian function $P(\omega)=(1/\sqrt{\pi}\sigma)e^{-\hbar(\omega-\omega_{\text{m}})^2/(2\sigma^2)}$, with center $E_{M}=\hbar \omega_{\text{m}}$ and disorder width $\sigma$. We define the range of the Gaussian distribution of excitation energies as $\Delta E$. The light field is quantized by the photonic operators $\hat{a}$ (under the single mode approximation). The light-matter interaction in $\hat{H}_{MC}$ is given by $g_{\text{c}} \cos(\Omega t + \phi)$, where $\Omega$ means the molecular rotational angular velocity and $\phi$ means the molecular rotational phase. When $2g_{\text{c}}\sqrt{N_{\text{M}}} \gg \Delta E$, there is a strong interaction between light and matter, and when $2g_{\text{c}}\sqrt{N_{\text{M}}} \ll \Delta E$, there is a weak interaction between light and matter. Due to the temporal periodicity of molecular rotations, we will use Floquet theory to treat time-dependent Hamiltonian $\hat{H}(t)$ as time-independent $\hat{H}^{\text{F}}$:
\begin{widetext}
\begin{equation}
\begin{aligned}
    \hat{H}^{\text{F}} =  \sum_{n,m=-N_{\text{F}}}^{N_{\text{F}}} (\hat{H}_0 + n\hbar \Omega \otimes \hat{I}) \delta_{nm} + \hat{H}_1 (\frac{1}{2} e^{\text{-i}\phi}\delta_{n,m+1} + \frac{1}{2} e^{\text{i}\phi} \delta_{n,m-1})
\end{aligned}
\end{equation}
\end{widetext}
where $\hat{H}_0$ means the diagonal part of $\hat{H}$ and $\hat{H}_1$ means the off-diagonal part of $\hat{H}$ removing $\cos(\Omega t + \phi)$. $\hat{I}$ is identity matrix. Dirac delta function $\delta$ is used. $N_{\text{F}}$ is the number of Floquet levels.

We restrict the current investigation to molecular rotational angular velocity and energetic disorder under a system of $N_{\text{M}} = 10^6$ molecules with $N_{\text{C}}=1$ cavity mode.

\textit{Sparse Floquet Hamiltonian.}—In dealing with such an extensive system, the primary obstacle shifts from computing speed to memory usage. To address this, and mindful of the sparsity within $\hat{H}^{\text{F}}$ whose dimension is [$(2N_{\text{F}}+1)(N_{\text{M}}+N_{\text{C}})$, $(2N_{\text{F}}+1)(N_{\text{M}}+N_{\text{C}})$], we opt for a memory-efficient approach. The entire matrix multiplication is founded on the non-zero elements using the Coordinate Format (COO) sparse matrix framework, identified as $\hat{H}_{s}^{\text{F}}$ whose total number of the non-zero elements is [$3 \times (2N_{\text{F}}+1)(N_{\text{M}}+N_{\text{C}})$] . This not only helps mitigate memory constraints but also streamlines the computational process.

\onecolumngrid
\twocolumngrid

\textit{Density of States (DOS).}—We use the implicitly restarted Lanczos method to compute half $(k/2)$ eigenvalues (denoted as $\Lambda$) and their corresponding eigenvectors (denoted as $U$) from each end of the spectrum of the sparse matrix $H_{s}^{\text{F}}$. This method is a wrapper for the ARPACK~\cite{arpack} SSEUPD and DSEUPD functions~\cite{lehoucq1998arpack}. Then the DOS $\rho$ of polaritons can be written as :
\begin{equation}
    \rho(\omega) = \text{Tr} [ \delta (\tilde{H}-\omega)U^{\dagger}P_{\text{C}}U ]
\end{equation}
where $\tilde{H} = \textbf{diag}(\Lambda)$ and $\textbf{diag}$ means diagonal matrix mapping. $P_{\text{C}}=\ket{\psi_{\text{C}}^{\text{F}}(0)}\bra{\psi_{\text{C}}^{\text{F}}(0)}$ is the projection operator to the cavity(photon) space and
\begin{equation}
    \psi_{\text{C}}^{\text{F}}(0) = \frac{1}{\sqrt{N_{\text{C}}}} \sum_{q=-N_{\text{F}}}^{N_{\text{F}}} \mathbf{1}_{[\text{{start}}(q) : \text{{end}}(q)]}
\label{eqn:psiF}
\end{equation}
where $\psi_{\text{C}}^{\text{F}}$ is an array with $N=(2N_{\text{F}}+1)(N_{\text{M}}+N_{\text{C}})$ elements, and $\mathbf{1}_{[\text{{start}}(q) : \text{{end}}(q)]}$ is an array of length $(N_{\text{M}}+N_{\text{C}})$, where all elements are zero except for the range $[\text{{start}}(q), \text{{end}}(q)]$, where the elements are set to $1$. The specific forms of $\text{{start}}(q)$ and $\text{{end}}(q)$ are given by:
\begin{equation}
\begin{aligned}
    \text{{start}}(q) =& (N_{\text{M}}+N_{\text{C}})(q+N_{\text{F}}), \\
    \text{{end}}(q) =& (N_{\text{M}}+N_{\text{C}})(q+N_{\text{F}}+N_{\text{C}}).
\end{aligned}
\end{equation}
Note that $\sum$ in Eq.~\ref{eqn:psiF} means array concatenation instead of numerical summation. Before taking the trace, the delta function is initially implemented numerically with Gaussian regularization.\cite{weisse2006kernel, bradbury2020stochastically, bradbury2023realistic} The Gaussian regularization can be interpreted as applying a Gaussian linebroadening to the traditional time correlation function. The density of states is then directly related to this transformed function by a specific transform:
\begin{equation}
\begin{aligned}
    \rho(\omega) &= \text{Tr}(\int dt \ e^{-\text{i}\tilde{H}t} e^{\text{i}\omega t} e^{\frac{\gamma^2t^2}{2}} \ U^{\dagger}P_{\text{C}}U) \\
    &=\text{Tr}(\frac{1}{\gamma \sqrt{\pi}} e^{-\frac{(\tilde{H}-\omega)^2}{\gamma^2}} \ U^{\dagger}P_{\text{C}}U)
\end{aligned}
\end{equation}
in the small $\gamma$ limit.
Through the transformation, we have reduced the originally calculated [$(2N_{\text{F}}+1)(N_{\text{M}}+N_{\text{C}})$, $(2N_{\text{F}}+1)(N_{\text{M}}+N_{\text{C}})$] dimension to a controllable [$k$, $k$] dimension, greatly reducing computational memory while not losing the key information we need about polaritons.

\onecolumngrid
\begin{center}
\begin{figure}
    \centering
    \includegraphics[width=.98\textwidth]{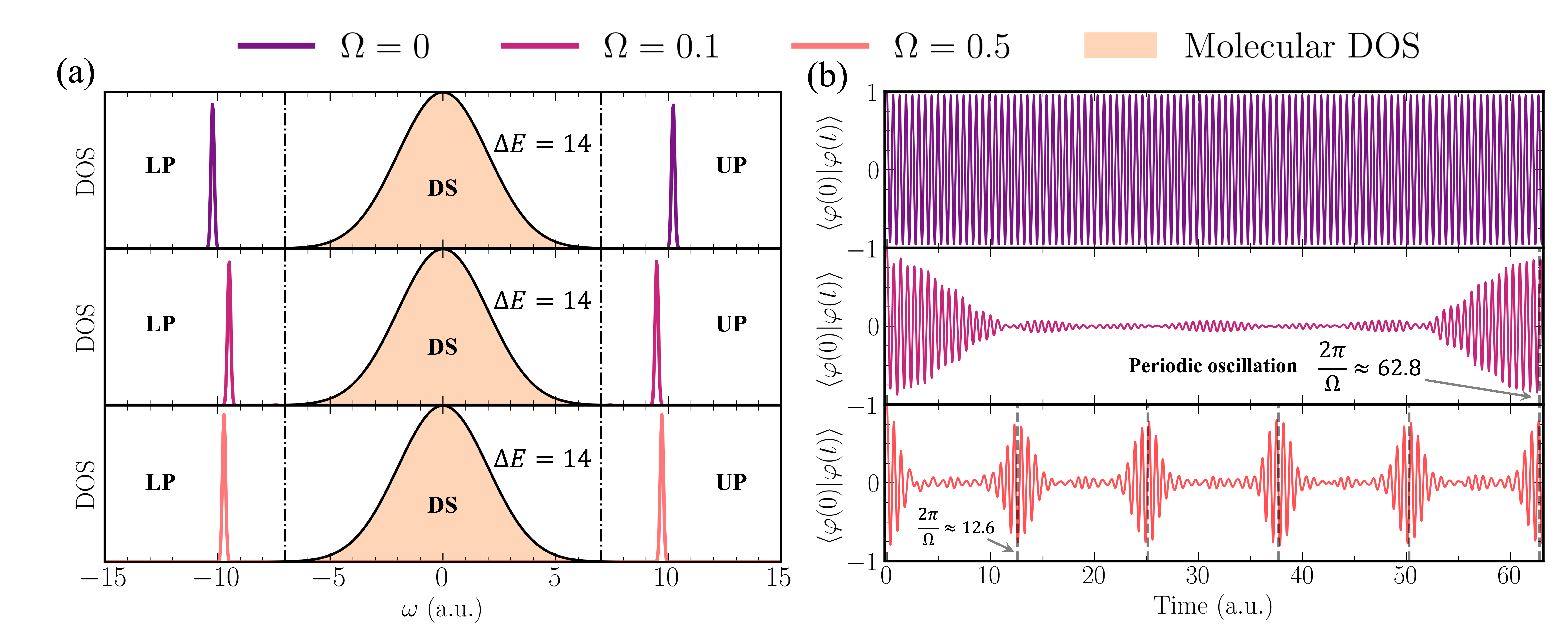}
    \caption{(a) Normalized density of states (DOS) with respect to the energy frequency $\omega$ under different molecular rotational angular velocity $\Omega$. The black dotted line represents the truncation position of the Gaussian distribution. $\Delta E$ represents the range of disordered molecular energy levels. Note that the amplitude of polariton and molecular DOS has been adjusted to display on the same graph. (b) Time evolution of correlation functions of photon-matter hybrid  wavefunctions $\varphi$ under different $\Omega$, as defined in Eq.~\ref{eqn:corr}. After introducing molecular rotation, the oscillation period of the photon-matter hybrid states will depend on the molecular rotation period ($\frac{2\pi}{\Omega}$, red dash lines). Parameters: $\Omega=0,0.1,0.5$, $\phi=0$, $k=30$, $\hbar=1$, $g_{\text{c}} = 0.01$, $\omega_{\text{c}}=\omega_{\text{m}}=0$, $\sigma=2$, $\gamma=0.1$, $N_{\text{F}}=3$, $\Delta t=0.001$, $N_{M}=10^{6}$, $N_{\text{C}}=1$.}
    \label{fig:omega}
\end{figure}
\end{center}
\twocolumngrid

\textit{Correlation Function.}—We calculate the correlation function in non-Floquet space, but at the same time, we need to use Floquet Hamiltonian to evolve the wavefunction, such as
\begin{equation}
    \text{i}\hbar \frac{\partial \psi_{\text{C}}^{\text{F}}(t)}{\partial t} = \hat{H}_{\text{s}}^{\text{F}} \psi_{\text{C}}^{\text{F}}(t).
\end{equation}
So we need a transformation from the wavefunction in Floquet space to the wavefunction in non-Floquet space.
\begin{equation}
\begin{aligned}
    &\varphi(t) = \\
    &\frac{1}{2N_{\text{F}}+1} \sum_{j=0}^{2N_{\text{F}}} \psi_{\text{C}}^{\text{F}}(t)[j(N_{\text{M}}+N_{\text{C}})]\exp\left(\text{i}\Omega(j-N_{\text{F}})t \right).
\end{aligned}
\end{equation}
Then the correlation function of photonic wavefunction will be
\begin{equation}
    \langle \varphi(0)|\varphi(t) \rangle = \varphi(0)^{\dagger}\varphi(t).
\label{eqn:corr}
\end{equation}

\textit{Analysis.}—In Fig.~\ref{fig:omega} (a), we explore the Density of States (DOS) of polaritons for varying parameters $\Omega=0, 0.1, 0.5$ with a strong light-matter interaction regime ($g_{\text{c}}=0.01$), considering $N_{\text{M}}=10^{6}$ molecules coupled to a single cavity mode. For $\Omega=0$ (no molecular rotation), the system exhibits a Rabi splitting of $\Omega_{Rabi} \approx 2g_{\text{c}} \sqrt{N_{\text{M}}}=10$, with $\omega_{\text{c}}=\omega_{\text{m}}=0$ leading to lower and upper polaritons. As $\Omega$ increases, the DOS shifts, indicating a dependency on molecular rotation. Particularly, at $\Omega=0.5$, the polariton DOS approaches that of the non-rotational case, revealing enhanced coupling strength compared to $\Omega=0.1$. The rotational motion induces inertial effects within molecules, altering the resonance frequency of polaritons. Molecular rotation influences the resonance frequency of molecular vibration, subsequently impacting polariton frequencies.

In Fig.~\ref{fig:omega} (b), the correlation function of the photonic wavefunction undergoes substantial changes for $\Omega=0, 0.1, 0.5$. For $\Omega=0$, the correlation function evolves uniformly and periodically. As $\Omega$ increases, the fluctuation period becomes dominated by molecular rotational angular velocity, with $T=2\pi / \Omega$. The change in fluctuation period from $\Omega=0.1$ to $\Omega=0.5$ highlights the profound impact of molecular rotation on photon state evolution. Molecular rotation induces changes in the spatial charge distribution, leading to a time-varying dipole moment that influences the amplitude and phase of polaritons.

\onecolumngrid
\begin{center}
\begin{figure}
    \centering
    \includegraphics[width=.98\textwidth]{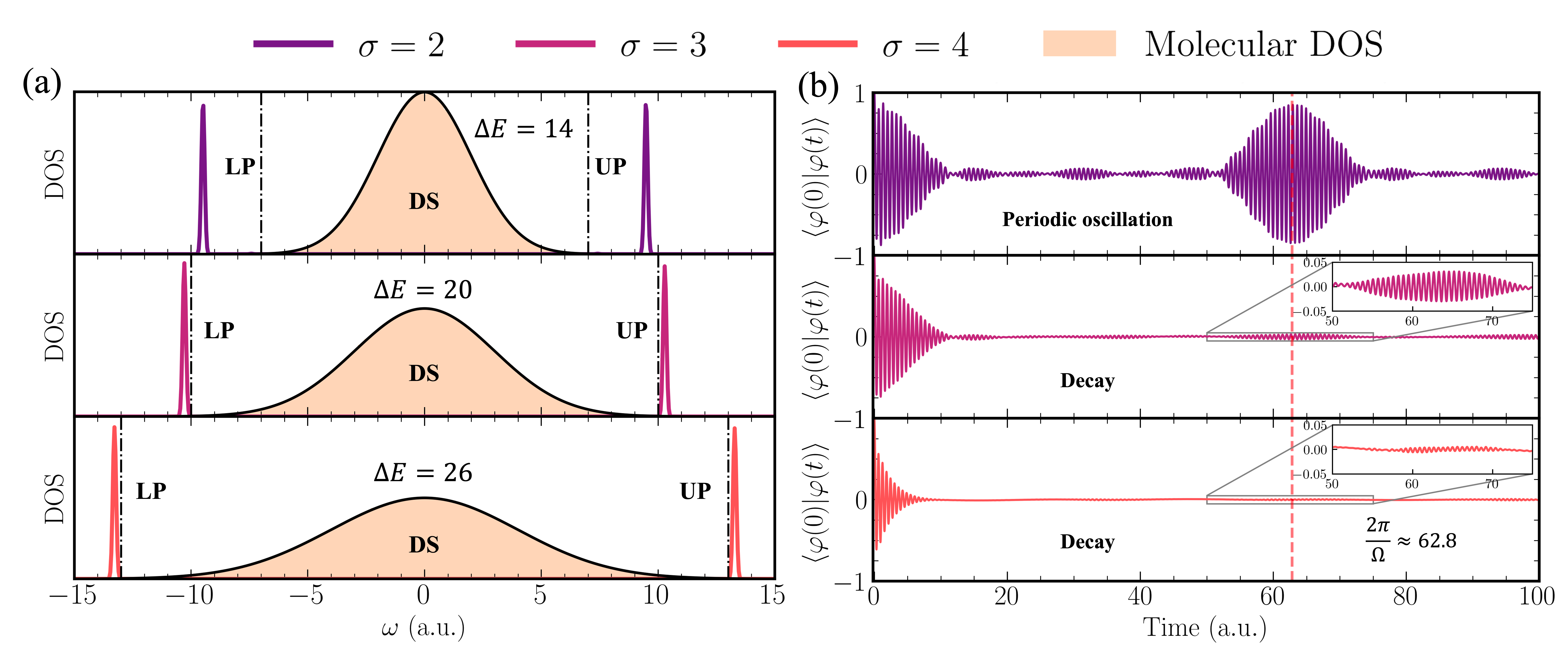}
    \caption{(a) Normalized density of states (DOS) with respect to the energy frequency $\omega$ under different disorder $\sigma$. The black dotted line represents the truncation position of the Gaussian distribution. As the disorder increases, the polaritons will become closer to the dark states (DS). $\Delta E$ represents the range of disordered molecular energy levels. Note that the amplitude of polariton and molecular DOS has been adjusted to display on the same graph. (b) Time evolution of correlation functions of photon-matter hybrid  wavefunctions $\varphi$ under different disorder $\sigma$, as defined in Eq.~\ref{eqn:corr}. After introducing molecular rotation, the oscillation period of the photon-matter hybrid states will depend on the molecular rotation period ($\frac{2\pi}{\Omega}$, red dash lines). When the disorder increases, the lifetime of polaritons rapidly decreases and approaches disappearance. Parameters: $\sigma=2,3,4$, $k=30$, $\hbar=1$, $g_{\text{c}} = 0.01$, $\omega_{\text{c}}=\omega_{\text{m}}=0$, $\Omega=0.1$, $\phi=0$, $\gamma=0.1$, $N_{\text{F}}=3$, $\Delta t=0.001$, $N_{M}=10^{6}$, $N_{\text{C}}=1$.}
    \label{fig:disorder}
\end{figure}
\end{center}
\twocolumngrid

In Fig.~\ref{fig:disorder} (a), we investigate the influence of molecular energy disorder (varying $\sigma$) on the polariton DOS. As $\sigma$ increases from $2$ to $4$, polariton energies deviate significantly from the center of molecular DOS. At the same time, these polaritons are very close to the dark states. Elevated disorder results in stronger spatial localization of polaritons, affecting their propagation and interaction within the optical cavity.

In Fig.~\ref{fig:disorder} (b), we analyze the impact of disorder $\sigma$ on the correlation function of photons. For $\sigma=3$ and $4$, the correlation function decays close to zero and loses its periodic fluctuations. Molecular energy level disorder broadens the energy levels, diminishing exciton lifetimes by increasing the transition rate between energy levels. This, in turn, affects the stability and duration of polaritons in the optical cavity. The interplay between energy level disorder and rotational motion introduces complexity to the excitation and relaxation processes of excitons, resulting in a dynamic behavior in the lifetime of polaritons.

\onecolumngrid
\begin{center}
\begin{figure}
    \centering
    \includegraphics[width=.98\textwidth]{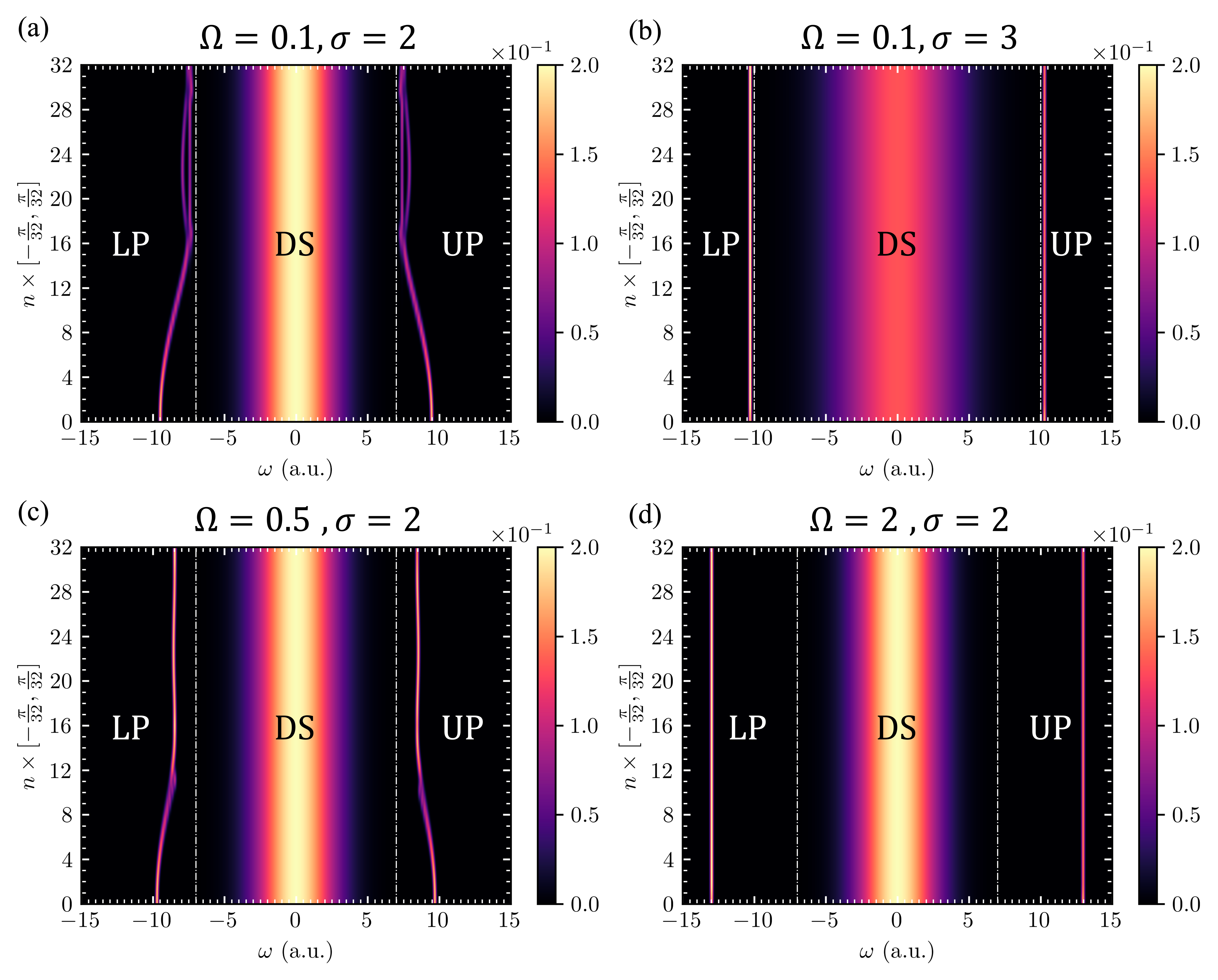}
    \caption{Heat maps of normalized density of states (DOS) with respect to the energy frequency $\omega$ under different phase $\phi$, which are sampled uniformly from $[-\frac{n\pi}{32},\frac{n\pi}{32}]$. The larger n, the greater the disorder of the phase. The white dotted line represents the truncation position of the Gaussian distribution.  Parameters: $k=30$, $\hbar=1$, $g_{\text{c}} = 0.01$, $\omega_{\text{c}}=\omega_{\text{m}}=0$, $\gamma=0.1$, $N_{\text{F}}=3$, $N_{M}=10^{6}$, $N_{\text{C}}=1$.}
    \label{fig:phase}
\end{figure}
\end{center}
\twocolumngrid

In Fig.~\ref{fig:phase}, we investigate the influence of molecular rotational phase disorder (varying $n$) on the polariton DOS. In Fig.~\ref{fig:phase} (a) and (c), we find that an increase in the disorder of rotational phase $\phi$ will cause the polariton DOS to approach DS when $\Omega=0.1$ or $0.5$, $\sigma=2$ and $\phi$ within the range from $0$ to $[-\frac{\pi}{2},\frac{\pi}{2}]$. Besides, double peaks will appear in the upper and lower polaritons when $\Omega=0.1$,$\sigma=2$, and $\phi$ within the range of $(-\frac{\pi}{2},\frac{\pi}{2})$ to $(-\pi,\pi)$. Increase $\Omega$ to $0.5$ and the bimodal pattern disappears. Further increasing $\Omega$ to $2$, UP and LP will significantly move to positions ($14$ and $-14$), but not only will UP and LP no longer approach DS as $n$ increases, but UP and LP are no longer symmetrically distributed, and the amplitude of UP is smaller than LP. As shown in Fig.~\ref{fig:phase} (b), we increase the $\sigma$ to $3$, and its Molecular DOS distribution range is within $[-10,10]$, which also produce the phenomenon shown in Fig.~\ref{fig:phase} (d). But at this point, UP and LP are very close to the edge of DS. From Fig.~\ref{fig:disorder} (b), we know that in this case, UP and LP will quickly degrade.

\begin{figure}
    \centering
    \includegraphics[width=.48\textwidth]{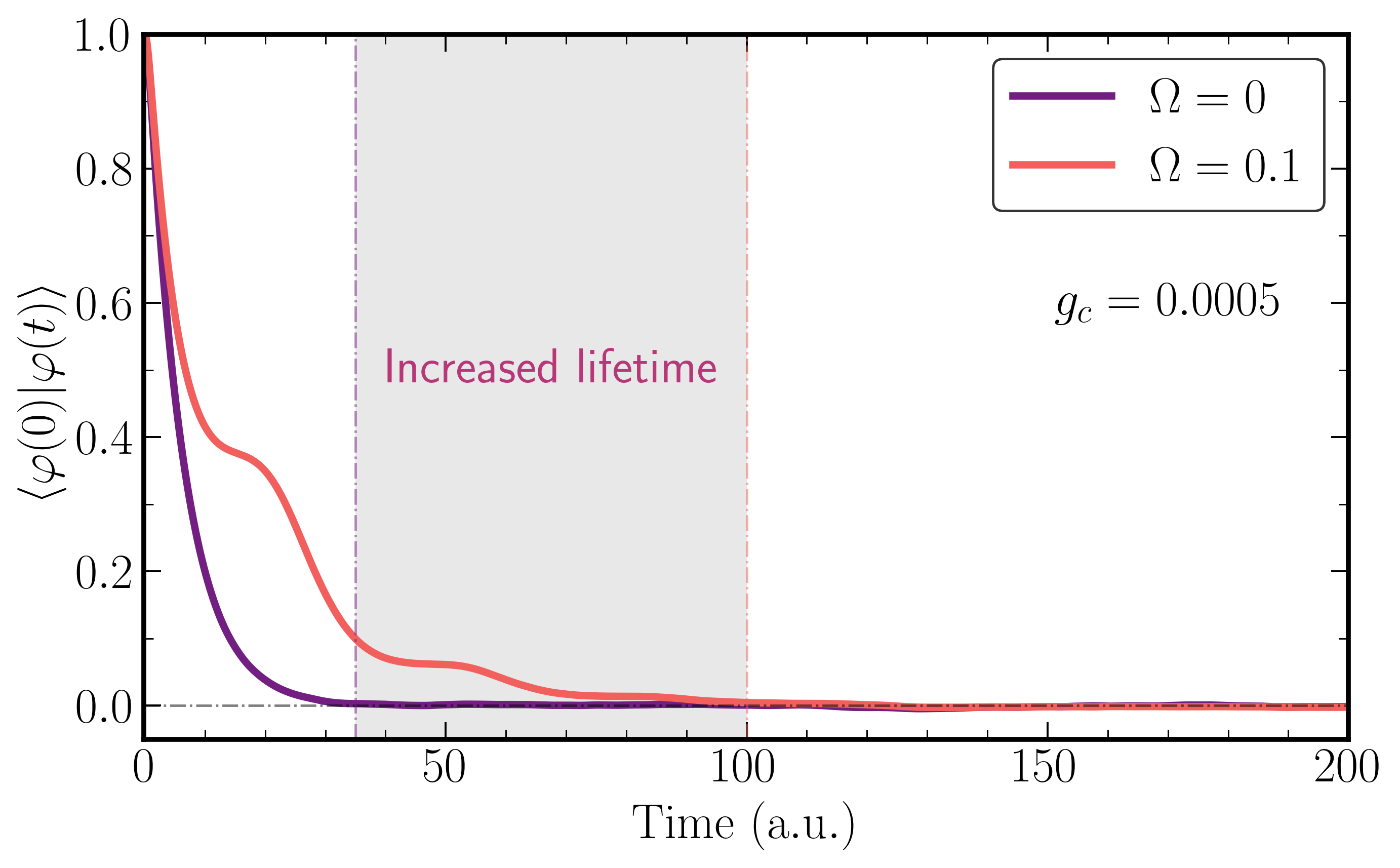}
    \caption{Time evolution of correlation functions of photonic wavefunctions $\varphi$ under weak coupling $g_{\text{c}}=0.0005$. We observe that molecular rotation increases the lifetime of the photon state. Parameters: $\Omega=0,0.1$, $k=30$, $\hbar=1$, $g_{\text{c}} = 0.0005$, $\omega_{\text{c}}=\omega_{\text{m}}=0$, $\sigma=2$, $\gamma=0.1$, $N_{\text{F}}=3$, $\Delta t=0.001$, $N_{M}=10^{6}$, $N_{\text{C}}=1$.}
    \label{fig:lifetime}
\end{figure}

Finally, in Fig.~\ref{fig:lifetime}, under weak light-matter interactions ($g_{\text{c}}=0.0005$), we observe that the introduction of molecular rotation enhances the lifetime of photon states.

\textit{Conclusion.}—Our investigation delves into the intricate interplay of molecular rotation and energy disorder in the context of strong and weak light-matter interactions. In the regime of strong interactions ($g_{\text{c}}=0.01$) illustrated in Fig.~\ref{fig:omega} (a), we find that the rotational motion of molecules significantly influences the DOS of polaritons. Notably, as the angular velocity $\Omega$ increases, a pronounced shift in the polariton DOS is observed, indicative of enhanced coupling strength. This effect is attributed to the inertial burden introduced by molecular rotation, impacting the resonance frequency of polaritons. Furthermore, the correlation function of the photonic wavefunction, as depicted in Fig.~\ref{fig:omega} (b), reveals a direct correlation between molecular rotation and temporal fluctuations in photon states. Molecular rotation induces changes in the charge distribution, impacting both the amplitude and phase of polaritons.

Turning our attention to disorder effects in Fig.~\ref{fig:disorder}, we analyze the impact of molecular energy disorder on polaritons. Increasing disorder ($\sigma=2$ to $\sigma=4$) leads to a substantial shift in polariton energies away from the molecular DOS center, indicating stronger spatial localization. This spatial confinement has profound implications for polariton propagation and interaction within the optical cavity. Correspondingly, disorder in molecular energy levels, as portrayed in Fig.~\ref{fig:disorder} (b), introduces energy level broadening, diminishing the lifetime of excitons. This reduction in lifetime is attributed to an increased transition rate between energy levels, affecting the stability and existence time of polaritons in the optical cavity. The combined influence of energy disorder and rotational motion results in a dynamic and complex interplay during the excitation and relaxation processes of excitons.

From a mathmatical perspective, our investigation demonstrates that the irregularity in molecular rotational phase (represented by the parameter $n$) exerts a significant impact on the DOS of polaritons. In Fig.~\ref{fig:phase} (a) and (c), it is observed that an increase in the irregularity of rotational phase $\phi$ leads to a gradual convergence of polariton DOS towards the DS when both the rotational velocity and disorder of the molecule are minimal, with $\phi$ varying within the range from $0$ to $[-\frac{\pi}{2},\frac{\pi}{2}]$. Additionally, a bimodal phenomenon is observed in the UP and LP when $\phi$ falls within the range of $(-\frac{\pi}{2},\frac{\pi}{2})$ to $(-\pi,\pi)$. However, when the rotational velocity of the molecule is substantial, the bimodal pattern disappears. At higher rotational velocities, the frequencies of the upper and lower polaritons become strongly correlated with the rotational speed of the molecule, no longer being influenced by the disorder in rotational phase. Simultaneously, the amplitude of the upper polariton consistently remains smaller than that of the lower polariton. 

In the context of weak light-matter interactions ($g_{\text{c}}=0.0005$), Fig.~\ref{fig:lifetime} demonstrates that the introduction of molecular rotation has a notable impact on the lifetime of photon states. The observed increase in photon state lifetime underscores the significance of molecular rotation even in the weak coupling regime.

In summary, our findings shed light on the intricate dynamics of polaritons influenced by molecular rotation and energy disorder. The presented results pave the way for a deeper understanding of these phenomena in the realm of cavity quantum electrodynamics and hold implications for the design and optimization of polaritonic devices. Future research in this direction could explore the potential applications of these insights in the design and optimization of optical cavities for specific functionalities, paving the way for advancements in the field of molecular-photon coupling.

This material is based upon work supported by National Natural Science Foundation of China (NSFC No. 22273075). W.L. acknowledges support from the high-performance computing center of Westlake University.

\nocite{*}

\bibliography{ref}

\end{document}